\documentclass[aps,prb,twocolumn,showpacs,preprintnumbers,amsmath,amssymb]{revtex4}

\usepackage[english]{babel}
\usepackage{graphicx}
\usepackage{subfigure}
\usepackage{dcolumn}
\usepackage{bm}
\usepackage{psfrag}

\begin{document}

\preprint{}

\title{Equation of state of two--dimensional $^3$He at zero temperature}

\author{M. Nava}
 \affiliation{Dipartimento di Fisica, Universit\`a degli Studi
              di Milano, via Celoria 16, 20133 Milano, Italy}
\author{E. Vitali}
\affiliation{IOM--CNR DEMOCRITOS National Simulation Center, via Beirut 2-4, 34014 Trieste, Italy}
 \affiliation{Dipartimento di Fisica, Universit\`a degli Studi
              di Milano, via Celoria 16, 20133 Milano, Italy}
\author{A. Motta}
 \affiliation{Dipartimento di Fisica, Universit\`a degli Studi
              di Milano, via Celoria 16, 20133 Milano, Italy}
\author{D.E. Galli}
 \affiliation{Dipartimento di Fisica, Universit\`a degli Studi
              di Milano, via Celoria 16, 20133 Milano, Italy}
\author{S. Moroni}
\affiliation{IOM--CNR DEMOCRITOS National Simulation Center, via Beirut 2-4, 34014 Trieste, Italy}
\date{\today}

\begin{abstract}
We have performed a Quantum Monte Carlo study of a two--dimensional bulk sample of interacting
$1/2$--spin structureless fermions, a model of $^3$He adsorbed on a variety of preplated 
graphite substrates.
We have computed the equation of state and the polarization energy
using both the standard fixed--node approximate
technique and a formally exact methodology, relying on bosonic imaginary--time correlation functions
of operators suitably chosen in order to extract fermionic energies.
As the density increases, the fixed--node approximation predicts a transition to an itinerant
ferromagnetic fluid, whereas the unbiased methodology indicates that the paramagnetic fluid is
the stable phase until crystallization takes place. 
We find that two--dimensional $^3$He at zero temperature crystallizes from the paramagnetic fluid at
a density of 0.061~\AA$^{-2}$ with a narrow coexistence region of about 0.002~\AA$^{-2}$.
Remarkably, the spin susceptibility turns out in very good agreement with experiments.
\end{abstract}
 

\maketitle

\section{INTRODUCTION}
A quasi--two--dimensional (2d) bulk $^3$He sample at zero temperature is
a very fascinating scenario to explore the physics of strongly correlated
Fermions. The liquid phase can be experimentally realized over a wide range of
densities by adsorbing $^3$He on a variety of preplated graphite substrates.
\cite{lusher,morhard,casey}
Heat capacity and magnetization measurements show that the system
displays a nearly perfect Fermi liquid behavior,
with the effective mass $m^{\star}$ and the enhancement of the spin 
susceptibility $\chi/\chi_0$ increasing with the density.
These observations, consistent with a divergence of $m^{\star}$ 
near the freezing density,
have been interpreted\cite{casey} as a signature of a Mott transition 
leading to an insulating crystal. However theoretical approaches 
\cite{krotscheck_mstar} suggest that the singularity of $m^{\star}$ 
and freezing could not have the same origin, and
indeed the freezing density (as well as the magnetic properties of
the solid\cite{fukuyama_review}, and even the presence of a possible 
intermediate phase of a uniform gas of vacancies 
\cite{fukuyama1}) is influenced by the preplated substrate. 
In order to characterize the sole effect of correlations, it is 
therefore of particular interest to study the ideal, strictly two--dimensional 
liquid on the verge of crystallization, in the absence of any external
potential. The measured properties of the fluid phase, on the other hand, 
appear to be largely independent of the substrate, so that they can be 
directly compared to the calculated properties of the ideal model 
\cite{whitlock}.

From the theoretical side, such a system provides a severe test 
case for microscopic calculations\cite{krotscheck}, because of 
the strong correlations attained at high densities.
We thus resort to quantum Monte Carlo (QMC) simulation,
a powerful tool to study strongly interacting systems, and
we calculate the ground--state energy per particle $e=\frac{E}{N}$
of the 2d $^3$He liquid at zero temperature
as a function of the number density $\rho$ and
the spin polarization $\zeta$.

The dependence of the energy on the spin polarization is in general very
weak in strongly correlated fluids\cite{ceperley_3d,holzmann,drummond,carleo}.
The so--called fixed--node (FN) approximation\cite{reynolds}, used in most 
QMC calculations, has been argued to give a significant bias in the 
polarization energy of three--dimensional liquid $^3$He\cite{holzmann} 
at high density. 

We thus perform our study going also beyond the
FN level, following a formally exact method\cite{carleo}, slightly
different from the well known transient estimate (TE) technique\cite{kalos}.
Briefly, we perform simulations relying on the basic Hamiltonian 
in an enlarged, unphysical space of states of any symmetry, including those
with Fermi and Bose statistics. The ground state 
energy of the physical fermionic $^3$He is considered as an excitation energy 
of the absolute bosonic ground state, which is sampled exactly with QMC. 
In this approach one trades the sign problem faced by TE\cite{kalos} for
the analytic continuation needed to extract excitation energies 
from suitable imaginary--time correlation functions. A mixed approach, 
devised to ease
detection of the asymptotic convergence of TE by a Bayesian analysis of 
imaginary--time correlation functions, was proposed By Caffarel 
and Ceperley\cite{Bagf}.

In fact a previous FN QMC calculation exists\cite{boronat}, but it is limited to 
low densities and only considers the paramagnetic fluid phase.
In particular, the accuracy of the FN approximation in the
high density regime is questionable\cite{holzmann}.
 
We find indeed that the FN level of the theory and the exact calculation
predict a qualitatively different behavior: within FN the system becomes
ferromagnetic well before crystallization takes place upon increasing
the density, whereas the unbiased calculation shows that the
spin polarization of the fluid is preempted by freezing, as observed
experimentally. From the estimated curve $e(\zeta)$ we obtain a
spin susceptibility enhancement in quantitative agreement with
the available measurements.

\section{QMC simulation}
We simulate $N$ particles with the mass $m_3$ of $^3$He atoms, interacting with
the HFDHE2 pair potential\cite{Aziz79} in periodic boundary conditions.
The simulation box, of area $\Omega$, is a square of side $L$ for the liquid phase;
for the solid it is a rectangle which accommodates a triangular lattice.
The Hamiltonian is

\begin{equation}
\label{hamiltonian}
\hat{H} = - \frac{\hbar^2}{2m_{3}}\sum_{i=1}^{N} \nabla_{i}^2
+ \sum_{i<j=1}^{N}v\left(\vec{\hat{r}}_i - \vec{\hat{r}}_j\right)
\end{equation} 

If the particles obey Bose statistics, projection QMC methods\cite{Reptation,pigs,Nava} provide 
unbiased estimates of the ground--state energy and other physical observables.
This is made possible by the formal similarity between Schr\"odinger equation
in imaginary time and the differential equation governing a diffusion process
in probability theory.
Such a useful possibility however drops down when Fermi statistics enters the 
game because of the well known sign problem\cite{kalos}.

\subsection{Fixed--Node approach}
The most commonly used approach in the QMC simulation of Fermions
is the fixed--node approximation\cite{reynolds}, which stochastically solves 
the imaginary--time Schr\"odinger equation  subject to the boundary conditions 
implied by the nodal structure of a given trial function $\Psi_T$. This approach 
gives a rigorous upper bound to the ground state energy, which often turns out
to be extremely accurate.

For a given spin polarization, i.e. considering $N_{\uparrow}$ spin--up and 
$N_{\downarrow} = N - N_{\uparrow}$ spin--down $^3$He atoms, $\Psi_T$
is chosen of the form
\begin{equation}
\Psi_T(\mathcal{R})= \mathcal{D}
(\mathcal{R}_{\uparrow})
\mathcal{D}
(\mathcal{R}_{\downarrow})
\Psi_J(\mathcal{R})\chi_{\zeta}
\end{equation}
where $\mathcal{R} \equiv (\vec{r}_1,...,\vec{r}_N)$,
$\mathcal{R}_{\uparrow} \equiv (\vec{r}_1,...,\vec{r}_{N_{\uparrow}})$,
$\mathcal{R}_{\downarrow} \equiv (\vec{r}_{N_{\uparrow}+1},...,\vec{r}_{N})$,
and the whole dependence on the spin degrees of freedom is contained
in $\chi_{\zeta}$,  a spin eigenfunction for the given polarization
\begin{equation}
\zeta = \frac{N_{\uparrow} - N_{\downarrow}}{N}\quad,
\end{equation}

The Jastrow factor,
\begin{equation}
\Psi_J(\mathcal{R})=\prod_{i<j}\exp\left(-\frac{1}{2}u\left(|\vec{r}_i-\vec{r}_j|\right)\right),
\end{equation}
describes pair correlations arising from the interaction potential;
we use a simple pseudopotential of the McMillan form $u(r)=(b/r)^5$. 
Finally, the simplest form of the antisymmetric factors $\mathcal{D}\left(\mathcal{R}_{\uparrow,\downarrow}\right)$
is in the form of Slater Determinants of plane waves:

\begin{equation}
\label{planewaves}
\mathcal{D}\left(\mathcal{R}_{\uparrow,\downarrow}\right)
= \det \left( \left\{\exp(i\vec{k}_i \cdot \vec{r}_j)\right\}_{i,j}\right)
\end{equation}
More accuracy in the FN results is achieved by introducing also backflow 
correlations\cite{Backflow} via quasi--particles vector positions:

\begin{eqnarray}
\label{backflow}
& \mathcal{D}\left(\mathcal{R}_{\uparrow,\downarrow}\right) = \det \left(
\left\{\exp(i\vec{k}_i \cdot \vec{x}_j)\right\}_{i,j}\right) \\ \nonumber
& \vec{x}_j \buildrel{def} \over {=} \vec{r}_j + \sum_{i\neq j = 1}^{N} \eta(|\vec{r}_j - \vec{r}_i|)
\left(\vec{r}_j - \vec{r}_i\right).
\end{eqnarray}
For the backflow correlation function $\eta(r)$ we adopt the simple form:

\begin{equation}
\eta(r)=A e^{-B(r-C)^2} \quad .
\end{equation}
We will refer to the two choices respectively as plane waves fixed--node (PW--FN)
and backflow fixed--node (BF--FN).
For each density, the variational parameters $b$, $A$, $B$ and $C$ are optimized
using correlated sampling\cite{rewate} at $\zeta=0$, and left unchanged at different
polarizations.

Part of the bias related to the finite size of the simulated system arises 
from shell effects in the filling of single--particle orbitals\cite{Lin}. 
This bias can be substantially reduced 
adopting twisted boundary conditions \cite{Lin}, i.e. choosing $\vec{k}$ 
appearing in \eqref{planewaves} and \eqref{backflow} inside the set:

\begin{equation}
\label{wavevectors}
\vec{k}_{\vec{n}} = \frac{2\pi\vec{n} + \vec{\theta}}{L}
\end{equation}
where $\vec{n}$ is an integer vector, $L$ is the side of the simulation box 
$\Omega$ and $\vec{\theta}$ is a {\it twist parameter} $\theta_i \in [0,\pi]$
which, at the end of the calculations, is averaged over. 

In the solid phase, quantum exchanges are strongly suppressed and the
energy difference between a Fermionic and a Bosonic crystal is
negligibly small for the purpose of locating the liquid--solid
transition. We will therefore replace the energy of $^3$He with that of
a fictitious bosonic Helium of mass $m_3$, which can be calculated exactly\cite{Reptation,pigs,spigs}.
The small error incurred by such replacement is bound by
the difference between the fermionic Fixed--Node (FN) energy and the unbiased bosonic energy.
This difference, calculated\cite{nosanow} as a check at the melting density where it is 
expected to be largest, is indeed in the sub--milliKelvin range.

\subsection{Fermionic correlations approach}
For the fluid phases the FN approximation may not be accurate enough,
particularly at high density where correlations are stronger and the energy
balance between different polarization states is more delicate.
Indeed, a FN study of three--dimensional $^3$He,
despite the use of sophisticated backflow correlations, strongly suggests
that this is the case\cite{holzmann}. We thus perform calculations beyond
the FN approximation, using a technique\cite{carleo} which is in principle 
exact, albeit practically limited to moderate system sizes.
The idea, in part related to
the transient estimate formalism\cite{kalos,Bagf}, is that 
of formally viewing \eqref{hamiltonian} as an
operator acting inside the Hilbert space $\mathcal{H}(N) \equiv \left(L^2(\Omega)\right)^{\otimes N}$,
that is {\it {forgetting}} spin and statistics: one can use Quantum Monte Carlo 
to sample the lowest energy eigenfuction $\psi_0(\mathcal{R})$ of $\hat{H}$
among the states of any symmetry.

It is known \cite{Ettore2} that $\psi_0$ must share the {\it {Bose symmetry}} of the Hamiltonian, so that:

\begin{equation}
\label{ebose}
E_0^B \equiv \frac{\langle \psi_0 | \hat{H} \psi_0 \rangle_{\mathcal{H}(N)}}{\langle \psi_0 | \psi_0 \rangle_{\mathcal{H}(N)}}
\end{equation}
is the Ground State energy of a fictitious system of $N$ Bosons of mass 
$m_{3}$ interacting via the potential $v(r)$.

The {\it bridge} that gives access to fermionic energies may be built up as follows: 
let us fix a spin polarization which is surely a good quantum number since the basic Hamiltonian is spin--independent. 
As discussed in Ref.\onlinecite{carleo}, if we are able to define an operator 
$\hat{\mathcal{A}}_{F}$ such that, inside $\mathcal{H}(N)$,

\begin{equation}
\psi_F\left(\mathcal{R}\right) = \left(\hat{\mathcal{A}}_{F}\psi_0\right)\left(\mathcal{R}\right) 
\end{equation}
has {\it non--zero overlap} with the configurational part of any {\it exact
fermionic} Ground State of $\hat{H}$ for the given $\zeta$, we can
define the {\it imaginary--time correlation function}:

\begin{equation}
\label{cfun}
\mathcal{C}_{F}(\tau)
 \equiv \frac{\langle \psi_0 | \left(e^{\tau \hat{H}}\hat{\mathcal{A}}_{F}^{\dagger}e^{-\tau \hat{H}}\right)
\hat{\mathcal{A}}_{F} \psi_0 \rangle_{\mathcal{H}(N)}}{\langle \psi_0 | \psi_0 \rangle_{\mathcal{H}(N)}}, \quad \tau \geq 0
\end{equation}
which can be straightforwardly evaluated in a bosonic QMC simulation\cite{Reptation,spigs,Nava}.
The lowest energy contribution in $\mathcal{C}_{F}(\tau)$ provides the {\it exact gap} 
between the fermionic and the bosonic ground states of the two--dimensional Fermi 
liquid; this can be readily seen by formally expressing \eqref{cfun} on the basis 
$\{\psi_n\}_{n \geq 0}$ of eigenvectors of $\hat{H}$ corresponding 
to the eigenvalues $\{E_n\}_{n \geq 0}$:

\begin{equation}
\label{cfun2}
\mathcal{C}_{F}(\tau)
= \sum_{n=0}^{+\infty}
e^{-\tau \left(E_n - E_0^B\right)}
\frac{|\langle \hat{\mathcal{A}}_{F}\psi_0 |
\psi_n \rangle|^2}{\langle \psi_0 | \psi_0 \rangle}
\end{equation}
A quite natural choice \cite{carleo} is to define $\hat{\mathcal{A}}_{F}$ borrowing
suggestions from the form of the trial wave function for 
the FN calculation, i.e.:

\begin{equation}
\label{operator}
\left(\hat{\mathcal{A}}_{F}\psi_0\right)\left(\mathcal{R}\right) \buildrel{def} \over {=}
\mathcal{D}
(\mathcal{R}_{\uparrow})
\mathcal{D}
(\mathcal{R}_{\downarrow})
\psi_0(\mathcal{R})
\end{equation}
where we can choose either the definition \eqref{planewaves} of $\mathcal{D}$  or 
the definition \eqref{backflow}. We will refer to such choices simply
as the plane waves fermionic correlations (PW--FC) and the backflow fermionic correlations (BF--FC).
Naturally the final results for the Bose--Fermi gap should 
coincide within statistical uncertainties, and the actual comparison
can be a good test for the robustness of the approach.

\section{Analytic Continuation}
Once we have achieved a QMC evaluation of $\mathcal{C}_{F}(\tau)$, the information 
about the Bose--Fermi gap $\Delta_{BF}= E_0 - E_0^B$ is contained in the
resulting correlation functions. The results for $\mathcal{C}_{F}(\tau)$ appear as 
simple smooth decreasing functions, whose values can be evaluated only in correspondence
with a finite number of imaginary--time values, say $\{\tau_0,\tau_1,\tau_2,...,\tau_l\}$;
moreover the data are perturbed by unavoidable statistical uncertainties.
The Bose--Fermi gap $\Delta_{BF}$ is thus hidden inside the sets of limited and noisy data. 
How can we extract it?
\begin{figure}[t]
\includegraphics[scale=0.45]{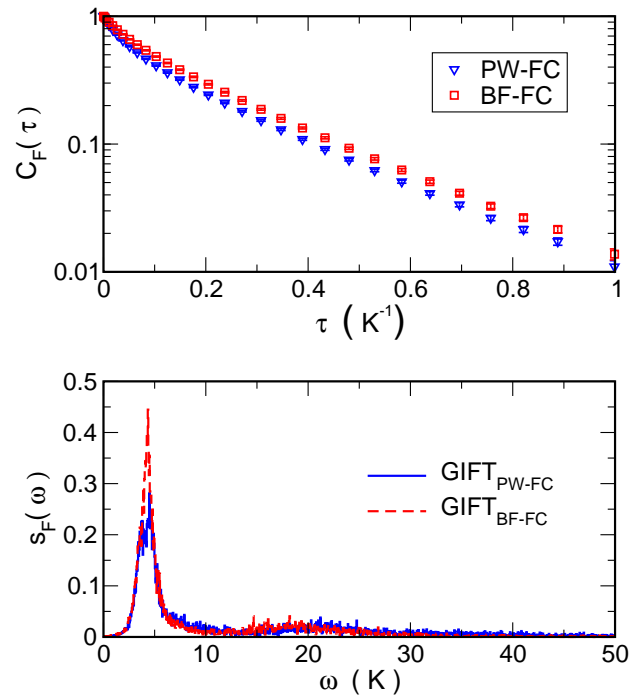}
\caption{(Color online) Upper panel: Imaginary time correlation functions, $\mathcal{C}_{F}(\tau)$,
corresponding to the two different choices of the operator in \eqref{operator}.
Lower panel: reconstructed spectral functions $s_F(\omega)$ obtained with the GIFT method. }
\label{pwbf}\end{figure}
In the upper panel of Fig.~\ref{pwbf} we show two imaginary time correlation functions
$\mathcal{C}_{F}(\tau)$, respectively a PW--FC and a BF--FC,
corresponding to the same spin polarization and twist parameter.
The long--$\tau$ tails of the two curves tend towards a
linear behavior (in logarithmic scale) with the same slope, and this
is a general feature shared by all the functions we have evaluated.
This indicates that, because of the finite--size of the system (and selection rules on the total momentum), the fermionic spectrum has a significant gap, i.e. the
lowest--energy term $\exp(-\Delta_{BF}\tau)$ in the correlation function \eqref{cfun2}
appears to be quite well resolved with respect to contributions from higher
fermionic energies.
The difference between the two curves (in particular the rigid shift between their
asymptotic tails) arises from the spectral weight
of the Ground State contribution, which is higher when backflow correlations
are taken into account, as expected.

In this favorable situation, the Bose--Fermi gap can be reliably extracted by 
simply fitting an exponential to the long--time tail of the correlation function.

This key result is strongly supported by a more sophisticated approach,
which evaluates $\Delta_{BF}$ by 
performing the full Laplace transform inversion of $\mathcal{C}_{F}(\tau)$,
i.e. solving 
\begin{equation}
\label{problem}
\mathcal{C}_{F}(\tau) = \int_{0}^{+\infty}d\omega e^{-\tau \omega} s_{F}(\omega) \quad ,
\end{equation}
for the unknown {\it spectral function} $s_{F}(\omega)$.
Recently a new method, the genetic inversion via falsification of theories (GIFT) 
method, \cite{Ettore} has been developed to face general inverse problems and in 
particular it has allowed to reconstruct the excitation spectrum of superfluid 
$^4$He starting from QMC evaluations of the intermediate scattering function 
in imaginary--time \cite{Ettore}; the results were in close agreement with 
experimental data \cite{Ettore}. Moreover the method has allowed to extract 
also multiphonon energies with a good accuracy level.
When applied to the two curves depicted in the upper panel of Fig.~\ref{pwbf},
we find the two spectral functions in the lower panel of Fig.~\ref{pwbf}; it is
apparent that the lowest--$\omega$ peak is indeed well resolved from higher--energy
contributions. Crucially, its position does not depend on the actual choice of the operator $\hat{\mathcal{A}}_F$, and it is in excellent agreement with the smallest
decay constant found by the simple exponential fit. 
The spectral weight instead is different, consistently with the differences
between PW--FC and BF--FC.

In this work we adopt an implementation of the inversion via falsification of 
theories, detailed in the Appendix, which avoids the rather CPU--intensive 
genetic algorithms\cite{goldberg}. This is crucial in the present study, which involves 
an extremely large number of reconstructed spectra. Indeed, 
a single QMC simulation for a given density produces
correlation functions pertaining to PW or BF operators, several 
spin polarizations, and 15 twist parameters in the irreducible wedge of the 
Brillouin zone of the simulation cell; on top of this, data are collected 
in several blocks, individually processed to obtain statistical uncertainties 
on the position of the lowest--energy peak.

\section{Results} 

\begin{figure}[t]
\includegraphics[scale=0.65]{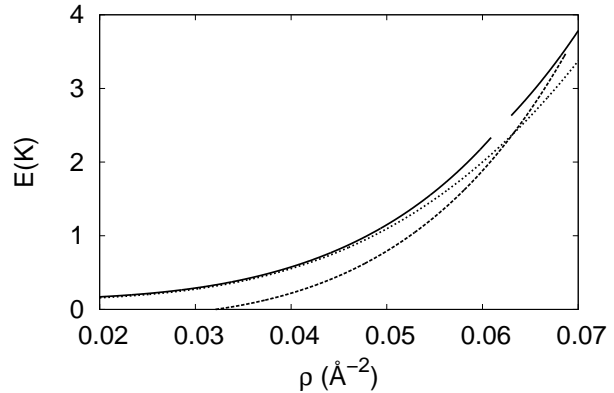}
\caption{Equation of state of $^3$He in two dimensions. Solid line (broken across the coexistence region): liquid and solid
         $^3$He; dashed line: mass--3 boson fluid; dotted line: liquid $^3$He, after Ref. \onlinecite{boronat}. The latter is
         only reliable at low densities.
         }
\label{stateq}
\end{figure}

We fit a fifth order polynomial to the density dependence of the energies of the triangular crystal and of the 
paramagnetic and the ferromagnetic fluids, listed in Table~\ref{tab1}.
\begin{table}[t]
 \caption{\label{tab1} The equations of state of $^3$He for the paramagnetic fluid and the solid
(solid lines in Figure~\ref{stateq})
are of the form $\alpha_1 \rho+\alpha_2 \rho^2+\alpha_3 \rho^3+\alpha_4 \rho^4+\alpha_5 \rho^5$.
This Table lists the values of the parameters $\alpha_i$, with lengths in \AA.
         }
  \begin{ruledtabular}
  \begin{tabular}{ccc}
  & liquid & solid \\
  \hline
$\alpha_1$ & 21.23782 & 57.35474 \\
$\alpha_2$ & -1344.413 & -2598.784 \\
$\alpha_3$ & 45093.37 & 58695.29 \\
$\alpha_4$ & -569306.0 & -532201.7 \\
$\alpha_5$ & 4383507  & 3063129 \\
  \end{tabular}
  \end{ruledtabular}
\end{table}
The resulting equation of state of two--dimensional $^3$He is shown in Figure~\ref{stateq}.
With the fermionic correlations method, we find a transition between the paramagnetic fluid 
and the triangular crystal around $\rho=0.061$~\AA$^{-2}$, with a narrow coexistence of about $0.002$~\AA$^{-2}$,
while the ferromagnetic fluid is never stable (see Table~\ref{tab2}).
\begin{table}[t]
 \caption{\label{tab2} Ground state energy of $^3$He in K, calculated by the FC method for the fluid phases and assumed
to equal the bosonic energy for the solid phase.
         }
  \begin{ruledtabular}
  \begin{tabular}{cccc}
  & liquid $\zeta=0$ & liquid $\zeta=1$ & solid \\
  \hline
0.020 & 0.1707(18) & 0.3218(25) &             \\
0.045 & 0.8168(86) & 0.9075(86) &             \\
0.050 & 1.1500(81) & 1.2123(88) &             \\
0.055 & 1.5972(93) & 1.6574(91) &             \\
0.060 & 2.2069(68) & 2.2493(54) & 2.2506(54)  \\
0.065 & 3.0065(73) & 3.0359(45) & 2.9195(26)  \\
0.070 & 4.0644(33) & 4.0915(34) & 3.7878(35)  \\
0.075 &            &            & 4.8728(44)  \\
0.080 &            &            & 6.2445(35)  \\
0.085 &            &            & 7.9589(39)  \\
0.090 &            &            & 10.0661(46) \\
0.095 &            &            & 12.6739(39) \\
0.100 &            &            & 15.8536(45) \\
  \end{tabular}
  \end{ruledtabular}
\end{table}
The energy of the bosonic mass--3 liquid is also shown. 
This fictitious system, simulated to extract the PW--FC and BF--FC energies, crystallizes at $\rho=0.069$~\AA$^{-2}$.
The freezing density of $^3$He is considerably higher than the highest density simulated in Ref.\onlinecite{boronat}. 
Correspondingly, the equation of state given in Ref.\onlinecite{boronat} is only reliable at relatively low density.
In particular, while it is only slightly below our results for $\rho\lesssim 0.045$~\AA$^{-2}$ as a consequence of the difference 
of interparticle potential adopted\cite{Aziz87}, it becomes (unphysically) even lower than the bosonic equation of state near the 
melting density, by an amount far larger than what could be due to the potential.

\begin{figure}[t]
\includegraphics[scale=0.45]{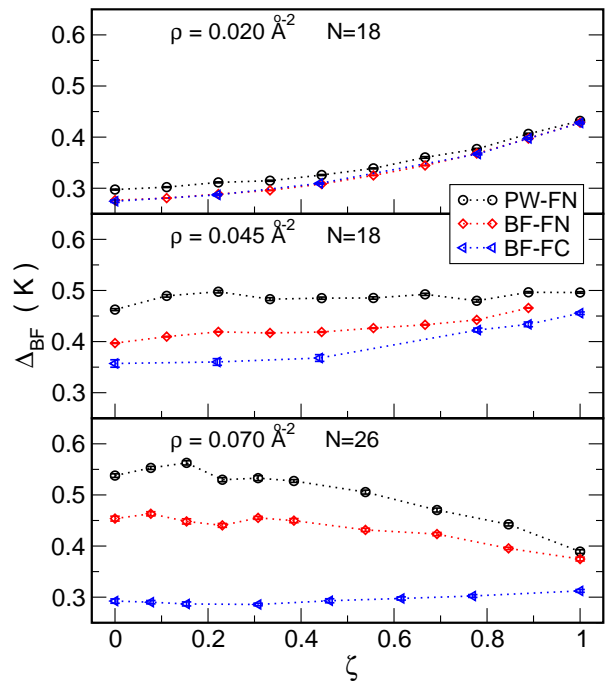}
\caption{(Color online) Upper panel: Bose--Fermi gap, $\Delta_{BF}$,
         as a function of the spin polarization, $\zeta$, at density $\rho=$0.020~\AA$^{-2}$
         evaluated via PW--FN, BF--FN, and BF--FC with $N=18$ particles.
         Middle panel: Bose--Fermi gap, $\Delta_{BF}$,
         as a function of the spin polarization, $\zeta$, at density $\rho=$0.045~\AA$^{-2}$
         evaluated via PW--FN, BF--FN, and BF--FC with $N=18$ particles.
         Lower panel: Bose--Fermi gap, $\Delta_{BF}$,
         as a function of the spin polarization, $\zeta$, at density $\rho=$0.070~\AA$^{-2}$
         evaluated via PW--FN, BF--FN, and BF--FC with $N=26$ particles.\\
         The statistical uncertainties are below the symbols size.}
\label{pcurves2}
\end{figure}

\begin{figure}[t]
\includegraphics[scale=0.65]{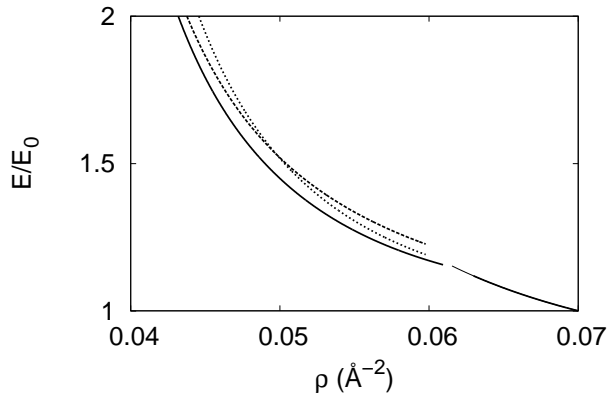}
\caption{Unbiased FC versus Fixed--Node equation of state. Thick solid line (broken across the coexistence region): paramagnetic
         liquid and solid $^3$He (FC); dashed line: paramagnetic liquid (FN); dotted line: ferromagnetic liquid (FN);
         the dashed and dotted lines terminate at the FN freezing density; thin solid line: energy of the solid, down to the
         FN melting density. For each density, the energy is relative to the energy $E_0$ of the mass--3 boson fluid.}
\label{stateq2}
\end{figure}

\begin{figure}[t]
\includegraphics[scale=0.65]{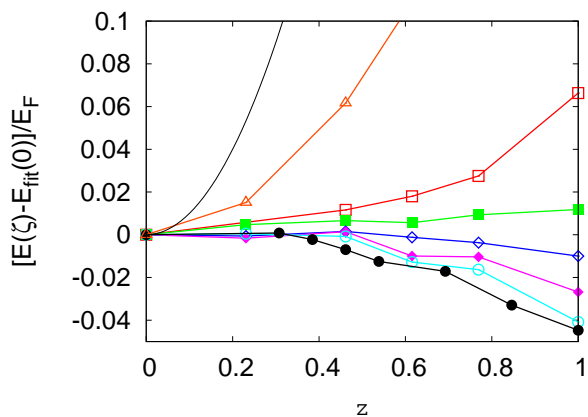}
\caption{(Color online) Fixed--node results for the polarization energy $E(\zeta)-E_{{\rm fit}}(0)$ relative to the Fermi energy
         $E_F$ at $\rho=0.020$ (open triangles), 0.045 (open squares), 0.050 (filled squares), 0.055 (open diamonds),
         0.060 (filled diamonds), 0.065 (open circles), 0.070 (filled circles)~\AA$^{-2}$, i.e. from top to bottom.
         The function $E_{{\rm fit}}(\zeta)$ is a quadratic polynomial in $\zeta^2$ fitted to the simulation data;
         the solid line is the density--independent result for non--interacting particles.}
\label{epol_fn}
\end{figure}

\begin{figure}[t]
\includegraphics[scale=0.65]{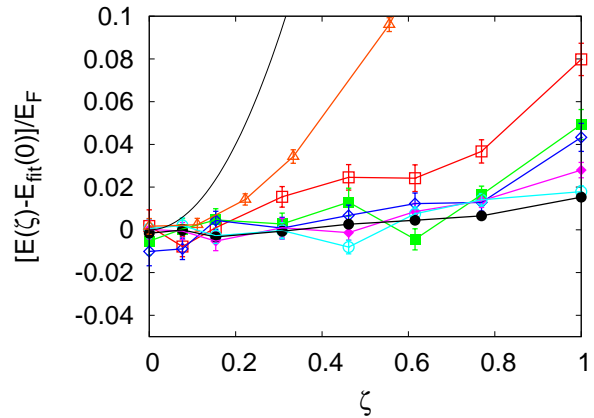}
\caption{(Color online) Exact results for the polarization energy $E(\zeta)-E_{{\rm fit}}(0)$ relative to the Fermi energy $E_F$ at
         $\rho=0.020$ (open triangles), 0.045 (open squares), 0.050 (filled squares), 0.055 (open diamonds),
         0.060 (filled diamonds), 0.065 (open circles), 0.070 (filled circles)~\AA$^{-2}$, in order of decreasing dispersion.
         The function $E_{{\rm fit}}(\zeta)$ is a quadratic polynomial in $\zeta^2$ fitted to the simulation data;
         the solid line is the density--independent result for non--interacting particles.}
\label{epol}
\end{figure}

The treatment of the spin polarization state requires a special care\cite{ceperley_3d,holzmann,drummond,carleo}.
In contrast to Ref.\onlinecite{boronat}, we find that the BF--FN energy can be significantly higher than the unbiased
Fermionic correlations (FC) energy. Starting from negligible values at low density,
the BF-FN error quickly increases approaching the strongly correlated
regime. As expected\cite{holzmann}, it is larger for the paramagnetic than for the ferromagnetic fluid. These
findings are exemplified in Figure~\ref{pcurves2}.
The inadequacy of the BF--FN is striking in the phase diagram: Figure~\ref{stateq2} shows that BF--FN incorrectly
predicts a transition to a ferromagnetic fluid well before crystallization takes place. 
Such a transition is also evident from Figure~\ref{epol_fn}, which shows the BF--FN results for the polarization
energy $e(\zeta)$ at various densities. The unbiased results, shown in Figure~\ref{epol}, display instead a paramagnetic
behavior even in a metastable fluid phase well beyond the freezing density.

\begin{figure}[t]
\includegraphics[scale=0.65]{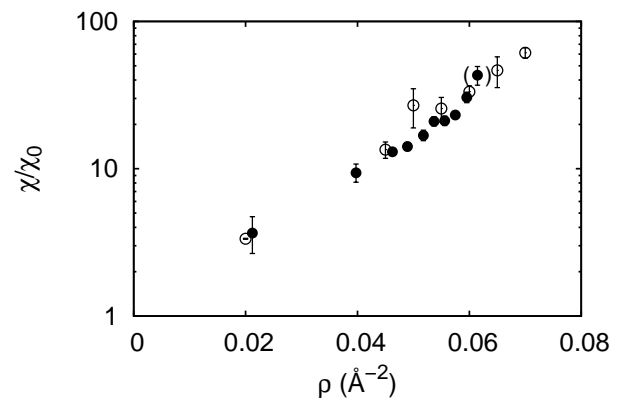}
\caption{Enhancement of the spin susceptibility as a function of the density:
         (filled circles) as measured in the second layer of $^3He$ on graphite\cite{morhard};
         (open circles) as calculated assuming a quadratic dispersion over the whole polarization range in Fig.~\ref{epol}.
         The corresponding Fixed--node result from Fig.~\ref{epol_fn} would diverge at $\rho\simeq 0.050$~\AA$^{-2}$.}
\label{chi}
\end{figure}

From the FC polarization energy $e(\zeta)$ we can estimate the spin susceptibility enhancement $\chi/\chi_0$. Assuming 
a quadratic dependence over the whole polarization range, which is generally consistent with the data
of Figure~\ref{epol}, we find an excellent agreement with the
measured susceptibility. Figure~\ref{chi} shows the comparison between the calculated $\chi/\chi_0$  
and the experimental data. We display only the results obtained in the second layer 
of $^3$He on graphite\cite{morhard} since they extend to the highest density in the 
fluid phase, but experiments carried on with differently preplated substrates
lead to equivalent results in their respective density ranges. The agreement among
the results obtained using different substrates induces us to expect that our ideal
model actually captures the physical mechanisms underlying the behavior of $\chi/\chi_0$.

\section{Conclusions}
We have calculated the equation of state and the polarization energy of 
$^3$He in two dimensions by means of an unbiased QMC method.
The system crystallizes into a triangular lattice from the paramagnetic fluid at 
a density of $0.061$~\AA$^{-2}$, with a narrow coexistence region of about $0.002$~\AA$^{-2}$; the ferromagnetic
fluid is never stable. From the polarization energy we obtain a spin susceptibility
enhancement in excellent agreement with the experimental values. 

The need for an exact QMC approach is witnessed by the failure of the Fixed Node
approximation with backflow correlations to predict the lack of a polarization
transition experimentally observed in the fluid phase, let alone an accurate value for
the spin susceptibility.

The estimation of the Bose--Fermi gap via the Fermionic correlation method is 
limited to relatively small systems: the present results are obtained with either 18 or
(in most cases) 26 particles. While the size effect remains the main source of
uncertainty of the present calculation, the agreement of the calculated and measured
spin susceptibility suggests that finite--size errors are relatively small.

\subsection{Acknowledgements}
This work has been supported by CASPUR, Regione Lombardia and CILEA Consortium through a LISA
Initiative (Laboratory for Interdisciplinary Advanced Simulation) 2010 grant [http://lisa.cilea.it] 

\appendix
\section{GIFT algorithm variant}
The inversion procedure that has been employed in this work is a variant of the GIFT algorithm\cite{Ettore}.
This new algorithm puts together the idea of 
the falsification principle\cite{Ettore} and a modified implementation of the
Prony's method\cite{Prony}, 
a non--iterative parametric technique for modeling using a linear combination of exponential functions.
Starting from the basic relation in \eqref{cfun2}, which has the general form
\begin{equation}
f(\tau)= \sum_{i=0}^{\infty} s_{i}e^{-\omega_{i}\tau} \quad ,
\end{equation}
provided that we are allowed to truncate the previous series, $\sum_{i=0}^{\infty} \leadsto \sum_{i=0}^{n-1}$,
the Prony's method is computationally very efficient (it runs in polynomial time) in deducing the coefficients
$\{s_{i}\}_{i=0}^{n-1}$ and $\{\omega_{i}\}_{i=0}^{n-1}$ from a limited set of estimations,
$f(k\delta\tau)=f^{*}_{k}$ at $k=0,1,\dots,2n-1$ of $f(\tau)$, being $\delta\tau$
the time step of the QMC simulation.
The main steps of this algorithm are the following:
\begin{enumerate}
\item solve the regularized linear system
\begin{equation}
\textbf{K}\textbf{a}=\textbf{b}
\end{equation}
defined by the Henkel matrix $K_{ij}= f^{*}_{i+j}$ and by the coefficients $b_{i}= f^{*}_{n+i}$ ($i,j<n$)
\item find the roots $\{z_{i}\}_{i=0}^{n-1}$ of the polynomial
\begin{equation}
z^{n}+a_{n-1}z^{n-1}+\dots+a_{1}z+a_{0}
\end{equation}
as eigenvalues of its respective companion matrix and obtain $\omega_{i}= -\frac{1}{\delta\tau}\ln z_{i}$
\item solve the regularized linear system
\begin{equation}
\textbf{A}\textbf{s}=\textbf{c}
\end{equation}
defined by the Vandermonde matrix $A_{ij}=z_{j}^{i}$ and by the coefficients 
$c_{i}= f^{*}_{i}$ ($i,j<n$)
\end{enumerate}
The transition from a nonlinear problem to two linear problems and one eigenvalues problem is the main
characteristic of this algorithm; from a mathematical and computational point of view this is an advantage.
Of course, the ill--posedness of this problem remains (our implementation uses the truncated singular value
decomposition regularization) and some care is
necessary to deal with instability against noise\cite{nmrprony}.
Such method fits the general scheme of the GIFT approach\cite{Ettore}, providing a very
efficient alternative to genetic algorithms in the
implementation of the falsification principle (when dealing with Laplace transform inversion).

\end{document}